\documentclass[preprint,12pt]{elsarticle}

\journal{Journal of Parallel and Distributed Computing}

\usepackage{subfigure}
\usepackage{graphicx}
\usepackage{amsmath}
\usepackage{amsthm}
\usepackage{amsfonts}
\usepackage{enumerate}
\usepackage[boxed,lined]{algorithm2e}

\begin{document}

\begin{frontmatter}




\title{
A GEMM interface and implementation on NVIDIA GPUs
for multiple small matrices
}


\author[]{Chetan Jhurani\corref{cor1}}
\ead{chetan.jhurani@gmail.com}
\cortext[cor1]{Corresponding Author}
\author[]{Paul Mullowney}
\ead{paulm@txcorp.com}

\address{
	Tech-X Corporation\\
	5621 Arapahoe Ave\\
	Boulder, Colorado 80303, U.S.A.
}


\begin{abstract}

We present an interface and an implementation of the General Matrix
Multiply (GEMM) routine for multiple small matrices processed
simultaneously on NVIDIA graphics processing units (GPUs). We focus on
matrix sizes under 16. The implementation can be
easily extended to larger sizes.  For single precision matrices, our
implementation is 30\% to 600\% faster than the batched cuBLAS implementation
distributed in the CUDA Toolkit 5.0 on NVIDIA Tesla K20c.  For example,
we obtain 104 GFlop/s and 216 GFlop/s when multiplying 100,000 independent
matrix pairs of size 10 and 16, respectively.  Similar
improvement in performance is obtained for other sizes, in single and
double precision for real and complex types, and when the number of matrices
is smaller.  Apart from our implementation, our different function interface also
plays an important role in the improved performance.  Applications of this
software include Finite Element computation on GPUs.

\end{abstract}


\end{frontmatter}



\newcommand{\matspace}[3]{\ensuremath{\mathbb{#1}^{#2 \times #3}} }

\newcommand{\Rnn}  {\matspace{\mathbb{R}}{n}{n}}
\newcommand{\Rmn}  {\matspace{\mathbb{R}}{m}{n}}
\newcommand{\Rmm}  {\matspace{\mathbb{R}}{m}{m}}

\newcommand{\Cnn}  {\matspace{\mathbb{C}}{n}{n}}
\newcommand{\Cmn}  {\matspace{\mathbb{C}}{m}{n}}
\newcommand{\Cmm}  {\matspace{\mathbb{C}}{m}{m}}
\newcommand{\Crn}  {\matspace{\mathbb{C}}{r}{n}}
\newcommand{\Crr}  {\matspace{\mathbb{C}}{r}{r}}


\newcommand{\vecspace}[2]{\ensuremath{\mathbb{#1}^{#2}} }

\newcommand{\Rn}   {\vecspace{\mathbb{R}}{n}}
\newcommand{\Cn}   {\vecspace{\mathbb{C}}{n}}

\newcommand{\Rm}   {\vecspace{\mathbb{R}}{m}}
\newcommand{\Cm}   {\vecspace{\mathbb{C}}{m}}

\newcommand{\reals} {\ensuremath{\mathbb{R}}}
\newcommand{\complex} {\ensuremath{\mathbb{C}}}


\newcommand{\norm}[1]{\ensuremath{\left| \left| #1 \right| \right|} }

\newcommand{\abs}[1]{\ensuremath{\left| #1 \right|} }

\newcommand{\pinv}[1]{\ensuremath{{#1}^{\dagger}} }
\newcommand{\inv}[1]{\ensuremath{{#1}^{-1}} }
\newcommand{\h}[1]{\ensuremath{{#1}^{op}} }

\newcommand{\nullsp}[1]{\ensuremath{\mathcal{N}(#1)} }
\newcommand{\rangsp}[1]{\ensuremath{\mathcal{R}(#1)} }

\newcommand{\patsym}[0]{\ensuremath{\mathcal{Z}} }
\newcommand{\pat}[1]{\ensuremath{\mathcal{Z}(#1)} }

\newcommand{\lagr}[0]{\ensuremath{\mathcal{L}} }

\newcommand{\partderiv}[2]{\ensuremath{\frac{\partial #1}{\partial #2}} }

\newcommand{\half}[0]{\ensuremath{\frac{1}{2}} }


\newcommand{\ordset}[3] {\ensuremath{\left\{ {#1}_{#2} \right\}_{#2 = 1}^{#3}}}


\newcommand{\Eq}[1]  {Equation~(\ref{#1})}
\newcommand{\Eqs}[1] {Equations~(\ref{#1})}
\newcommand{\Sec}[1] {Section~\ref{#1}}
\newcommand{\Fig}[1] {Figure~\ref{#1}}
\newcommand{\Alg}[1] {Algorithm~\ref{#1}}
\newcommand{\Rem}[1] {Remark~\ref{#1}}
\newcommand{\Prop}[1] {Property~\ref{#1}}
\newcommand{\Thm}[1] {Theorem~\ref{#1}}
\newcommand{\Def}[1] {Definition~\ref{#1}}
\newcommand{\Tab}[1] {Table~\ref{#1}}


\newtheorem{theorem}{Theorem}[section]
\newtheorem{lemma}[theorem]{Lemma}
\newtheorem{proposition}[theorem]{Proposition}
\newtheorem{corollary}[theorem]{Corollary}
\newtheorem{conjecture}[theorem]{Conjecture}
   
\theoremstyle{definition}
\newtheorem{definition}[theorem]{Definition}
\newtheorem{example}[theorem]{Example}
\newtheorem{examples}[theorem]{Examples}

\theoremstyle{remark}
\newtheorem{remark}[theorem]{Remark}



\section{Introduction}

Implementations of the General Matrix Multiply (GEMM) routine typically achieve a
large fraction of peak speed on modern multi-core hardware.  However,
because of hardware characteristics, high performance is achieved for
large matrix sizes, which is usually in hundreds or even thousands~\cite{fermi_gemm,Goto},
or for sizes that are integer multiples of hardware-specific dimensions.

Although such large dense matrices are important in many applications and also help
in showcasing the impressive speeds, many applications require
multiplication of multiple independent small matrices, each having identical
size.  Such a situation is common in finite element methods for
example~\cite{spectralHP,deville2002hom,cite:hpbook,cite:hpbook2,solin2003higher,uhm}.
Even if sizes are different, one can partition matrices
so that sets with identical matrix sizes are obtained. 

Our goal is to obtain high performance when processing multiple small
matrices on NVIDIA GPUs, numbering in thousands or much higher.  The
definition of small is not standard and cannot be invariant with
respect to hardware trends. Still, `small' can be taken to be a size
for which high performance is difficult to achieve in a parallel
environment unless multiple independent matrices are involved.
Note that even for large matrices, efficient implementations of GEMM
multiply sub-blocks which are smaller and achieve high performance.
However, the difference here is that our small matrices are independent
and are of different sizes.

Achieving high performance GEMM for small matrix sizes, when compared to large
sizes, is inherently difficult because each entry is used fewer times after
it is copied from main memory to registers.  However,
developing a high-quality GEMM implementation is crucial.  Apart from
its inherent utility, fast GEMM is also a basis for the speeds
achievable in other level-3 BLAS functions~\cite{cite:gemmblas}.

In the context of GPUs, the need for a capability to multiply pairs of small matrices
was recognized by NVIDIA. An
implementation focused on small matrix sizes was released with the
Compute Unified Device Architecture (CUDA) toolkit version 4.1 in the
cuBLAS library.  The so-called ``batched'' implementation (function name
{\tt
cublasXgemmBatched, X = S/D/C/Z} for different data types) is
significantly faster than two other possibilities -- one using CUDA
streams and the original non-streamed version intended for large
matrices.  This improvement in speed due to batching holds for matrix sizes up to
roughly 100.  However, except for small matrices with sizes that
are integral multiples of 16, the achieved performance is still a small
fraction of peak GPU speed~\cite{cublas}. One can also observe that the function's
performance is quite sensitive to the matrix size being a multiple of
16.  For example, its speed in multiplying 100,000 single precision 
matrix pairs of size 16 is 134 GFlop/s. But it drops to 105 GFlops/ for size 15, and for
size 17 it drops even more to 32 GFlop/s.  These numbers are using the ``batchCUBLAS''
example program run on Tesla K20c, a device for which the peak single precision speed is 3.52
TFlop/s.

Another issue is that the function interface of {\tt cublasXgemmBatched} is
designed around pointers to the matrix pointers, which increases its
applicability when calling from languages that can deal with pointers
to pointers and if different matrices were allocated separately. But this
feature also makes it less efficient because one has to transfer each
pointer as extra data.

Our objectives in this contribution are to achieve higher
performance GEMM for small matrices on GPUs and also to provide an
alternative interface that uses a
second-level leading dimension (a BLAS/LAPACK terminology~\cite{cite:lapack}).
Using our interface and the implementation, we achieve an improvement
in speed that varies in [30\%,600\%] compared to the reference cuBLAS
batched implementation for single precision matrices.  The improvement is in the higher range when the matrix sizes
are smaller. Similar improvements are observed for other scalar types.

If we just use the cuBLAS-like interface but the underlying
implementation is ours, the speedup is lower but still
significant.   In this case, compared to the reference, we achieve a more modest speed improvement (or
a minor reduction in a tiny fraction of cases) in [-30\%,300\%].  We also include discussion
of a few C++ template related features that help in achieving generality
and high performance.

Our focus here is on square matrices of size 16 or less but the method can be
easily implemented for rectangular cases and for larger sizes.  The
main reason we have focused our attention on matrices of size 16 and
less is that such sizes correspond to the polynomial degrees that
are traditionally used in
high-order finite elements~\cite{spectralHP,deville2002hom,cite:hpbook,cite:hpbook2,solin2003higher,uhm}.

Here is an outline of the paper.  In~\Sec{sec:gemm} we briefly
describe our notation for batched GEMM.  Sections~\ref{sec:cublas_intf}
and~\ref{sec:tradeoffs} are devoted to describing the cuBLAS interface
that we use as reference and the design trade-offs associated with it.
In Sections~\ref{sec:lda2} and ~\ref{sec:cpp} we describe the interface
of our CUDA kernel and the function wrapping it.  There we also describe
various optimization possible due to C++ features.  We give
an overview of our implementation in~\Sec{sec:impl}.  We discuss
the performance of our implementation from various viewpoints in
\Sec{sec:gemm_result}.

\section{Multiple GEMMs}
\label{sec:gemm}

Let $op$ (standing for operation) refer to a mapping from matrices to
matrices, such that $op(A) = A$ or $A^T$ or $A^*$ depending on an extra
variable that can contain three values.  The superscripts
$T$ and $*$ stand for transpose and Hermitian-transpose, respectively.
Consider real or complex matrices $A^p, B^p, C^p$ for $p =
1,2,\ldots,N$.  Here $N$ stands for the batch size.  The matrices can
be stored in single precision or double precision.  Using the BLAS
notation, $C$ is $m \times n$, $op(A)$ is $m \times k$, and $op(B)$ is
$k \times n$.

Let $\alpha$ and $\beta$ be given scalars. Our goal is to compute
\begin{equation}
\label{eq:gemm}
C^p \leftarrow \alpha \, op(A^p) op(B^p) + \beta \, C^p
\end{equation}
for $p = 1,2,\ldots,N$ independently in the CUDA environment.
Note that in this context the operation $op$ acting on $A$ can be
different than the one acting on $B$, which leads to 9 combinations in
all.

\section{NVIDIA cuBLAS interface}
\label{sec:cublas_intf}

We first provide a few details of our baseline.  The batched
implementation in NVIDIA cuBLAS versions 4.1, 4.2, and 5.0 for computing the
output in~\Eq{eq:gemm} is a C function with a signature given below.
The letter $T$ (standing for type) refers to the concrete type, which
can be {\tt S,D,C,Z} for {\tt float}, {\tt double}, {\tt cuComplex},
or {\tt cuDoubleComplex} respectively.  The detailed documentation is
available online~\cite{cublas}.  However, the interface is easy
to understand and almost natural for a BLAS/LAPACK user.
The comments after {\tt //} are our own.
\begin{verbatim}
cublasStatus_t cublasTgemmBatched(
  cublasHandle_t handle,    // library context
  cublasOperation_t transa, // CUBLAS_OP_[NTC] for A, A^T, A^*
  cublasOperation_t transb, // CUBLAS_OP_[NTC] for B, B^T, B^*
  int m, int n, int k,      // matrix sizes, discussed earlier
  const T *alpha,    // host or device pointer
  const T *Aarray[], // device pointers to 0,0 elements of A's
  int lda,           // leading dimension of A's
  const T *Barray[], // device pointers to 0,0 elements of B's
  int ldb,           // leading dimension of B's
  const T *beta,     // host or device pointer to beta
  T *Carray[],       // device pointers to 0,0 elements of C's
  int ldc,           // leading dimension of C's
  int batchCount);   // number of A's, B's, and C's
\end{verbatim}

\section{Design trade-offs in the cuBLAS interface}
\label{sec:tradeoffs}

The cuBLAS interface is natural when using multiple matrices that may
have been created and stored independently or if the number of
matrices is dynamically changing.  However, the interface has a few
design issues.

First, one would have to collect the pointers to the
matrices and copy the three arrays (for $A,B,C$) to the device.
Although natural for independently allocated matrices, this leads to
additional overhead in transmission as well as memory access during
kernel execution.  Secondly, when batch size is large, say a few
thousands or more, one would reduce the effective performance by
spending time in allocating and deallocating a large number of independent matrices and
transferring the arrays of pointers.

An alternative way to use the cuBLAS interface is to allocate a large chunk of memory once, and
store pointers to appropriate positions so that it looks like a 3-D
array -- a uniformly-offset collection of matrices of identical size.
This strategy will avoid number of allocations and deallocations to grow
with $N$, the batch size.
However, even now, neither the function invocation nor the kernel execution
have used the knowledge that the matrices
are uniformly-offset.  This leads us to the hypothesis that a somewhat
less general but a more efficient as well as appropriate way is to not
pass pointers to pointers, but use a second leading dimension to
indicate the offset between adjacent matrices.  One needs to pass
pointer to the 0,0 element of only the first matrix and just a single
extra integer for another leading dimension.

To test this hypothesis, we will show results from two versions of our kernel, one with the
interface like the one in cuBLAS and other with the 3-D array
interface.  In both cases, the implementation is be identical as far as
logically possible. It will be shown that the second one increases performance
even after transmission overhead associated with creating and passing
pointers to pointers is ignored (see ~\Sec{sec:gemm_result}). 
Hence, we can conclude the following based on our argument and evidence.  In the regime of
large batch size and small matrix sizes (less than or equal to 16), the natural
interface is the one that uses a 3-D array to represent a collection of
2-D matrices.

\section{A specialized function interface}
\label{sec:lda2}

\begin{figure}
\centering
\includegraphics[scale=0.5]{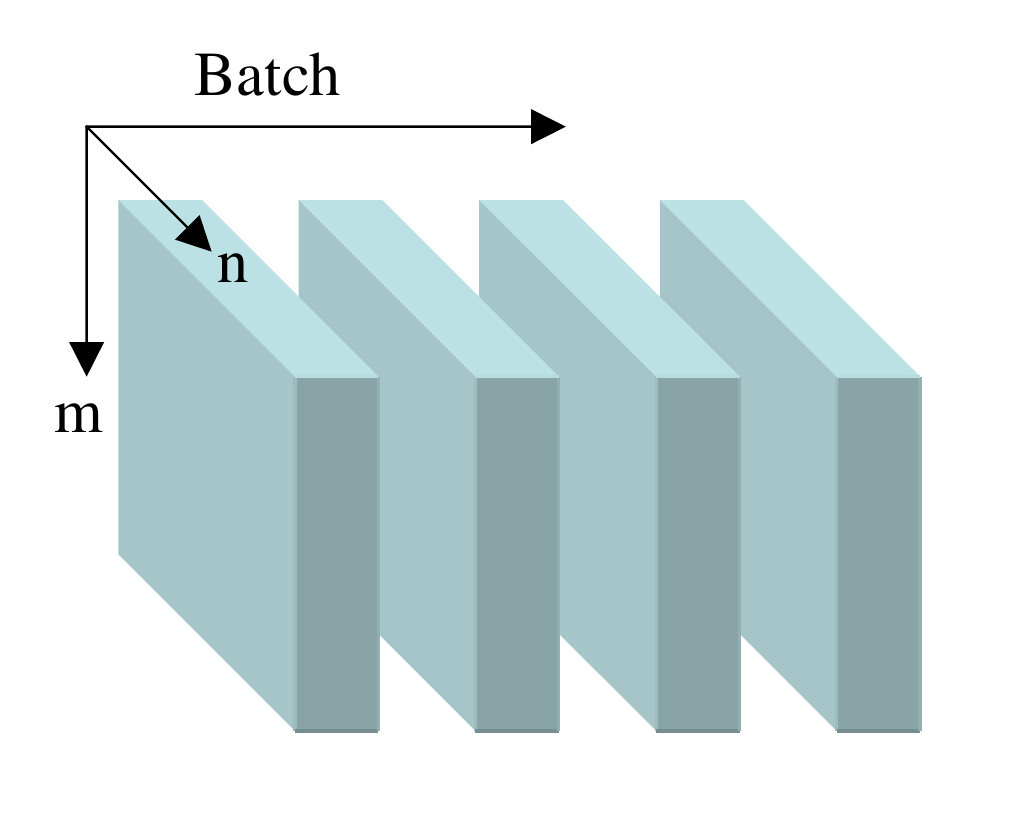}
\caption{A layout of matrices in the {\tt TGEMM\_multi\_uniform} kernel.  The dimensions
$m$ and $n$ refer to the output matrices $C^p$.  We have $ldc \geq m$ and $ldc2 \geq ldc * n$.
}
\label{fig:LD2}
\end{figure}

As mentioned earlier, we develop kernels with two kinds of interfaces
to test which interface gives us the better performance.  The first
one resembles the cuBLAS interface.  We call it {\tt TGEMM\_multi\_nounif},
where `multi' stands for multiple and `nounif' is for `not uniform'.
The second one is called {\tt
TGEMM\_multi\_uniform} and is the focus of this research. It uses a
second-level leading dimension.  Hence we show the second and new
interface only.
\begin{verbatim}
cudaError_t TGEMM_multi_uniform(
  char transa,       // [nN], [tT], [cC] for A, A^T, A^*
  char transb,       // [nN], [tT], [cC] for B, B^T, B^*
  int m, int n, int k, // matrix sizes, discussed earlier
  const T *alpha,    // host or device pointer
  const T *A3D,      // device pointer to 0,0,0 element of A's
  int lda,           // leading dimension of each 2D A
  int lda2,          // offset between 2D A's
  const T *B3D,      // device pointer to 0,0,0 element of B's
  int ldb,           // leading dimension of each 2D B
  int ldb2,          // offset between 2D B's
  const T *beta,     // host or device pointer to beta
  T *C3D,            // device pointer to 0,0,0 element of C's
  int ldc,           // leading dimension of each 2D C
  int ldc2,          // offset between 2D C's
  int batchCount);   // number of A's, B's, and C's
\end{verbatim}
The second leading dimension is the offset between adjacent 2D
matrices.  The $(i,j)$ entry in $k^{th}$ $A$ matrix in the batch
is $A[i + lda*j + lda2*k]$.  Similar relations hold for $B$ and $C$.
\Fig{fig:LD2} shows the layout of the batch of $C$ matrices.

The ``physical unit'' of leading dimension is an important issue
when writing code using CUDA and can be a cause of bugs if mishandled. The
units of leading dimensions, both the original one and new one, are not
bytes.  The units are number of scalar elements.  This has to be translated
when interfacing with CUDA functions that work with bytes. 

\section{A C++ kernel interface}
\label{sec:cpp}

We now present the C++ interface for the CUDA kernel underlying the
{\tt TGEMM\_multi\_uniform} function.  It is intended for
multiplication of square matrices.  Because of templates, it can be
called by C++ code only.  However, the C++ caller code can be wrapped
by passing appropriate arguments so it can be called by a C function
in the usual manner.
\begin{verbatim}
template
<
    typename T,         // single/double real/complex
    unsigned int m,     // size m = n = k
    bool transa,        // true if transposing A, else false
    bool transb,        // true if transposing B, else false
    typename unary_a,   // functor for conjugating A or not
    typename unary_b,   // functor for conjugating B or not
    typename axpby_type // functor for y <- a*x + b*y
>
__global__ void TGEMM_multi_uniform_kernel(
    const T* A,                 // A data
    unsigned int lda, int lda2, // A LDA 1 and 2
    const T* B,                 // B data
    unsigned int ldb, int ldb2, // B LDA 1 and 2
          T* C,                 // C data
    unsigned int ldc, int ldc2, // C LDA 1 and 2
    unsigned int batch_count,   // Number of A,B,C
    unary_a func_a,   // functor for conjugating A or not
    unary_b func_b,   // functor for conjugating B or not
    axpby_type axpby) // functor for y <- a*x + b*y
\end{verbatim}

A C++ functor (or function object) is useful for compile-time polymorphism~\cite{Stroustrup:2000:CPL:518791}
and is important for achieving high-performance.  The essential idea is to
move Boolean checks from the innermost loop to outside so it will lead
to fewer branching possibilities.  The drawback is that it leads to
larger executable code size when a large number of independent
template permutations are instantiated.  Of course, if only one or few
instantiations are used, then one gets high performance without
increasing code size significantly.

Note that our intent is not at all to suggest that cuBLAS or other
libraries do not internally use functors or templates like these.  Based on the
cuBLAS interface, the only knowledge we have is that it uses pointers
to matrices rather than second-order leading dimension.  It may still
be designed using such functors or templates internally.

We briefly discuss the specific rationale and a few candidates for the
{\tt axpby\_type} and {\tt unary\_a} argument types.  These functors
are applied entry-wise to pairs of scalars.  In many cases, the
GEMM parameters $\alpha$ (for $a$) and $\beta$ (for $b$) are not arbitrary
scalars but are simple values like -1, 0 or 1.  In such cases explicit
multiplication is not needed.  For example, one may just want to
compute $C \leftarrow A B$ (so $\alpha = 1, \beta = 0$).  The effect on
performance can be noticeable large when $\beta$ is zero but a general code
still reads the output entry from global memory, multiplies it by $\beta$,
and then writes the final sum back to global memory.  Such extra
computation and memory movement can be prevented by passing a stateless object
of the following type.
\begin{verbatim}
template<typename T>
struct axpby_a1b0 // For a = 1, b = 0
{
    __host__ __device__
    void operator()(T& y, const T& x) const
    {
        y = x;
    }
};
\end{verbatim}
The following functor can be used for general $a$ and $b$.
\begin{verbatim}
template<typename T>
struct axpby // For general a and b
{
    T a, b;
    axpby(const T& a, const T& b)
      : a(a), b(b)
    {}
    __host__ __device__
    void operator()(T& y, const T& x) const
    {
        y = a * x + b * y;
    }
};
\end{verbatim}
One would then call {\tt axpby(C\_entry, (AB)\_entry)} in the CUDA
kernel for each entry.  Functors for other special combinations of $a$
and $b$ can also be used.  Note that these functors are reusable for
other BLAS routines, on GPUs and CPUs alike, and are a minor coding
overhead at worst and significant performance boost in general.
Compiler optimizations inline all these calls.

In a similar vein, when one wants to apply conjugates for complex
scalars, functors can be used that return the conjugate of a given
entry.  Here is an example for double precision complex.
\begin{verbatim}
struct conjugate_cuDoubleComplex
{
    __host__ __device__
    cuDoubleComplex operator()(const cuDoubleComplex& a) const
    {
        cuDoubleComplex a_conj = {a.x, -a.y};
        return a_conj;
    }
};
\end{verbatim}
The following functor is used when conjugate is not needed.
\begin{verbatim}
template<typename T>
struct identity // for identity transform
{
    __host__ __device__
    T operator()(const T& a) const
    {
        return a;
    }
};
\end{verbatim}

\section{CUDA kernel implementation overview}
\label{sec:impl}

We give a description of the {\tt TGEMM\_multi\_uniform\_kernel}
implementation.  In reality, we have two distinct methods.  The first
one is for matrix sizes 1-16. The second can be used for
matrices where the square matrix dimension can be factorized
into a product of two nearly equal numbers.  For example, 15 can
be factorized as $3 \times 5$ or $5 \times 3$.  The order is
important.  Experimentally, we saw that the second method
is faster than the first one for sizes 15 and 16 and we use that
to show the results. In both methods, each CUDA thread-block is
used to process multiple matrices.

In the first method, each thread within a thread-block computes one entry of
output matrix $C$.  All the threads read the corresponding $A$ and
$B$ matrix to shared memory.  Thus, for processing $m \times m$
matrices we launch $m^2$ threads. 

In the second implementation, each thread can be used to read
multiple entries of $A$ and $B$ into the shared memory.  This
kernel is inspired from the one implemented in MAGMA~\cite{fermi_gemm}.
We factorize $m = m_1 m_2$ and launch thread-blocks of size
$(m_1 m_2) \times m_2$.  Then each block has to read multiple columns
of $A$ and $B$.

We use 2000 thread-blocks.  This
parameter can be tuned for minor additional gains for specific
situations (see~\Fig{fig:num_mat}). The gains will not be large unless the hardware is quite
different.  For matrices that are not square,
one can use these square matrix kernels to multiply sub-blocks.

Another important point to note is that one can design more specialized
kernels of a given matrix size so that the performance can be
improved even more.  However, we have not pursued this task further.

\section{Performance results}
\label{sec:gemm_result}

Our test hardware is a Tesla K20c GPU, which has a peak performance
of 3.52 TFlop/s and 1.17 GFlop/s in singe and double precision,
respectively.  We work with the ECC (error-correcting code) mode
turned off for all cases. Our code has been compiled with {\tt -arch=sm\_35 -O3}
options using the CUDA 5.0 toolkit.

Each experiment is designed to understand the performance variation
when a set of parameters is changed.  Thus, it is important to
remember the basic set of parameters around which the variations are
tested.  This is our basic set of parameters.
\begin{itemize}
\item We use 100,000 as batch size.
\item We use 2000 thread-blocks.
\item We use our kernel that takes second-order leading dimension parameter.
\item The no transpose or conjugate-transpose operation is our baseline.
\item The memory layout of matrices is such that all the leading dimension parameters
are the smallest they can be for the given matrix size.
\end{itemize}
We clearly mention when some of these are changed in an experiment.
The choices above have been made just to design
concrete experiments. They are not a restriction in any way.

In our results, the suffix `unif' is used to specify the leading dimension based
kernel, the suffix `nounif' is used for our implementation with cuBLAS like interface,
and the suffix `cuBLAS' is used for the cuBLAS implementation.

We have used the convention that $2m^3$ flops are required to
perform GEMM on two real $m \times m$ matrices, for all $\alpha$ and $\beta$
and whether transpose or conjugate-transpose is chosen or not.
For complex matrices, we use $8m^3$.
The ``3M'' method is not used
for complex matrix multiplication~\cite{higham_3m}.
For timing purposes, each kernel run was performed 5 times and we took
the median of the time reported by CUDA timers after the device was
synchronized.

\subsection{Block size and batch size}

Our first goal is to determine a block size for large number of
matrices such that the performance is close to maximum.
We do this for general values of $\alpha$ and
$\beta$.

\begin{figure}
\centering
\includegraphics[page=1,scale=0.4]{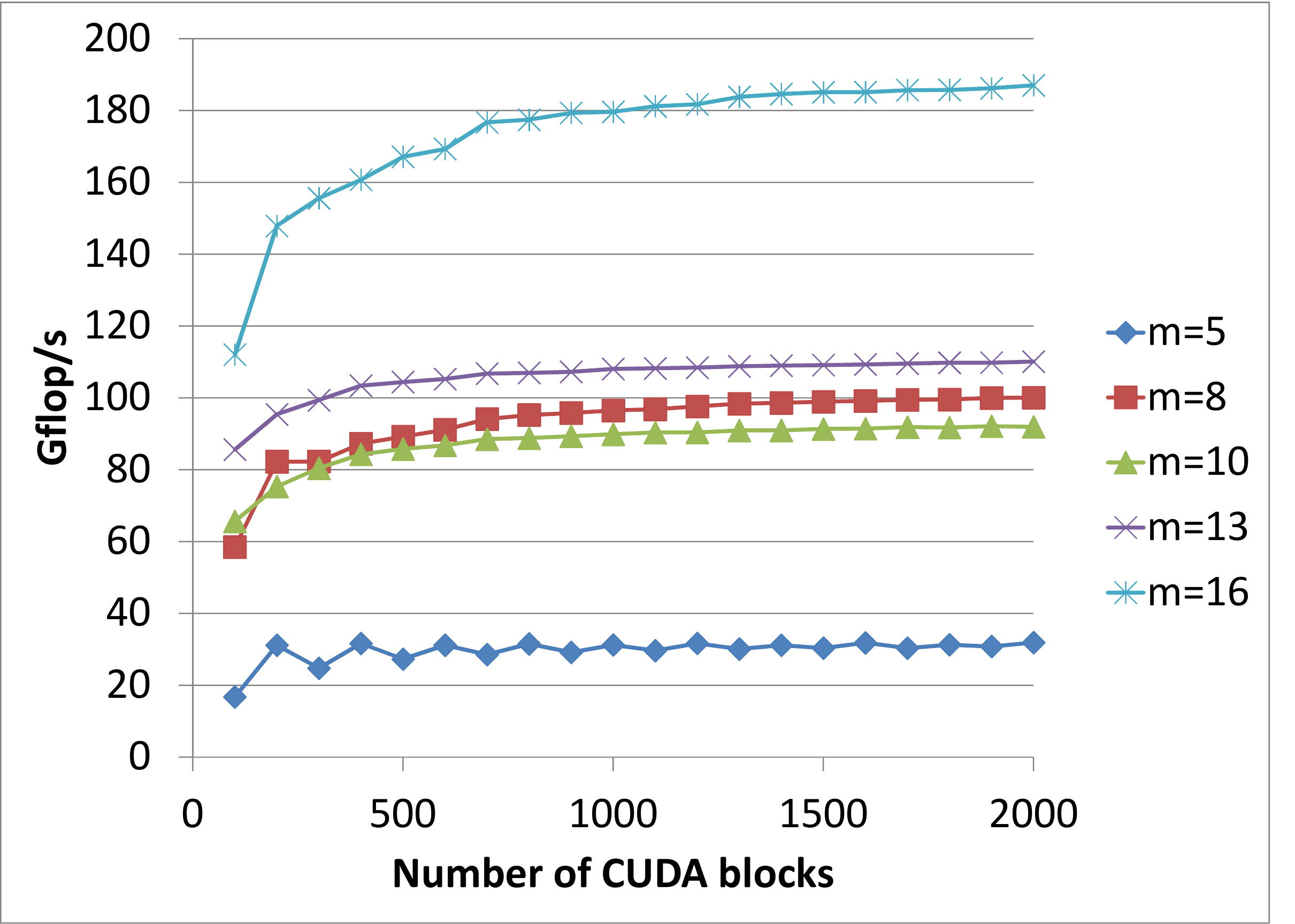}
\caption{An experiment to determine the effect of CUDA block size. We multiply 100,000
independent single-precision square matrix pairs of different sizes ($m$).}
\label{fig:block_size}
\end{figure}

\begin{figure}
\centering
\includegraphics[page=2,scale=0.4]{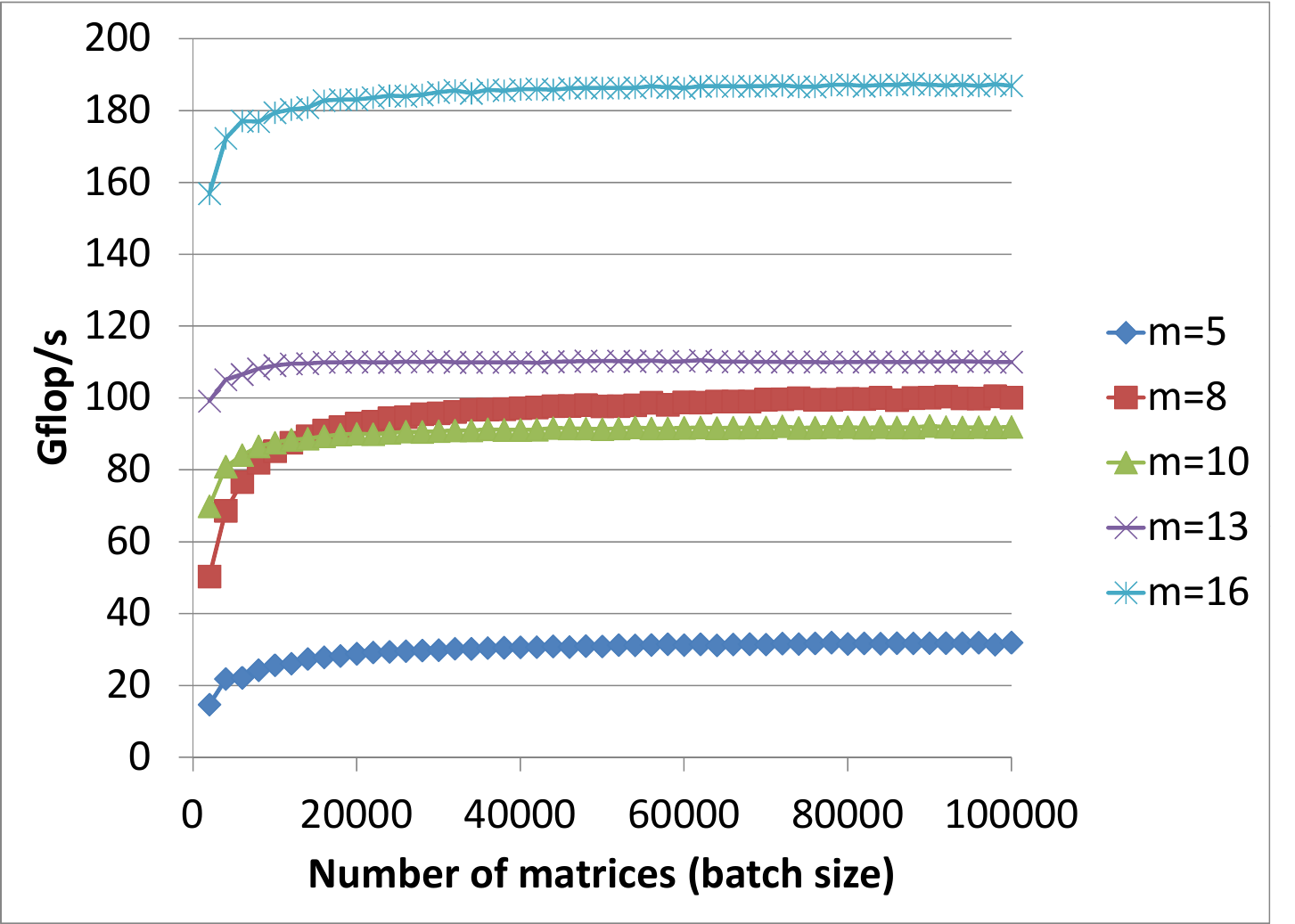}
\caption{An experiment to show that the performance improvement
increases very slowly with increasing batch size for various matrix sizes $m$.
100,000 matrix pairs in a batch is a reasonable parameter to run
other experiments.}
\label{fig:num_mat}
\end{figure}

\Fig{fig:block_size} shows the speed achieved when multiplying $N =
100,000$ independent single-precision square matrix pairs of different
sizes. Clearly, there is very little to gain when running a kernel
with more than 2000 blocks.  We chose $N$ matrices in a batch to stay
well within the limits of GPU memory.  For example, in the largest
case of double precision complex matrices, one would consume roughly
1.23 GB in storing 100,000 $A$,$B$, and $C$ matrices.  Another reason
for choosing that number is shown in~\Fig{fig:num_mat}.  There is some
speed gain when using large batch sizes but as one can extrapolate,
the gains would be minor beyond the threshold of $N = 100,000$.  We
fixed the block size and batch size based on these two experiments.
Another result from the second experiment is that even for small batch
sizes, say 2000, the performance is not significantly worse compared
to large batch sizes.

\subsection{Effect of conjugation or transposition on performance}

\begin{figure}
\centering
\includegraphics[scale=0.4]{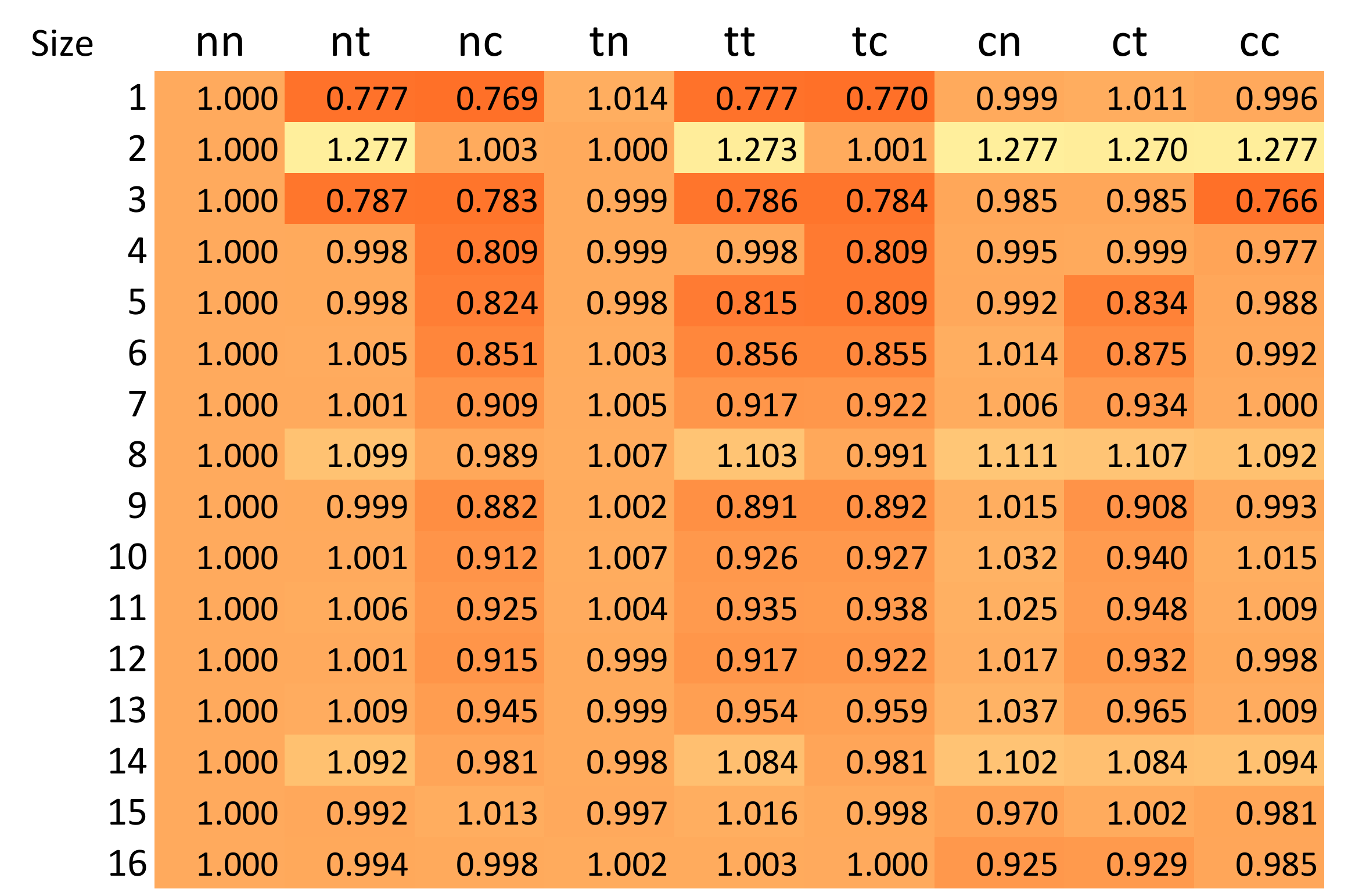}
\caption{Performance effect of conjugating (c) and/or transposing (t)
one or both sets of input single precision complex matrices with the
baseline being no operation (n). Slower performance is indicated by
darker colors and lower values (less than 1).}
\label{fig:conj_trans}
\end{figure}

In all our other experiments we have not conjugated and/or transposed
the matrices that are inputs to GEMM. \Fig{fig:conj_trans} shows that
these operations do make our kernels slower, which is
expected, but the reduction in speed is minor.
The results are shown for single precision complex matrices of size up
to 16.  This experiment was run for general (randomized) values of $\alpha$ and $\beta$.

\subsection{Improvement due to axpby\_type template parameter}

\begin{figure}
\centering
\includegraphics[page=3,scale=0.4]{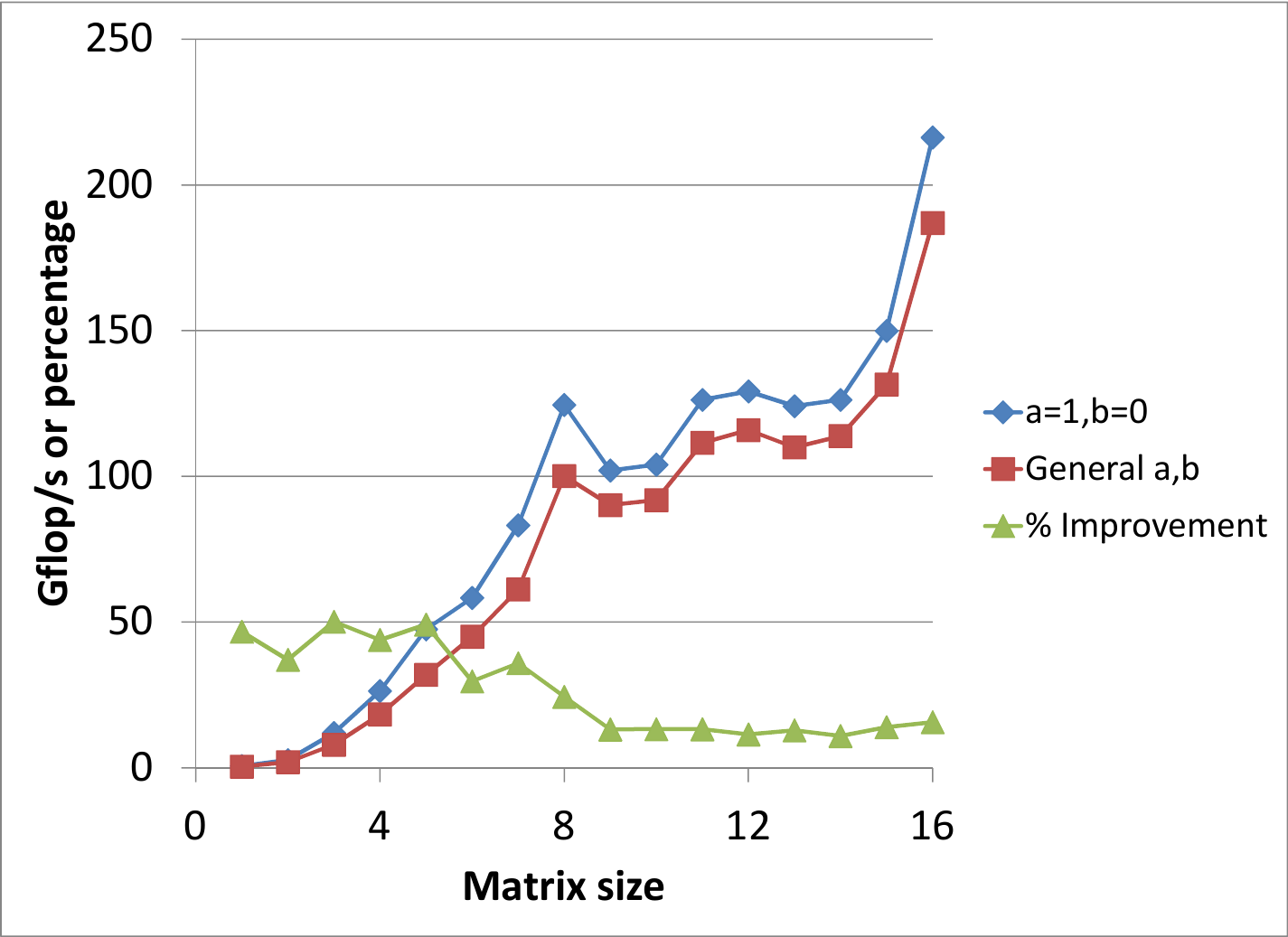}
\caption{Speed and performance gain for
single precision matrices of various sizes
when $\alpha = 1$ and $\beta = 0$ and for general values
of these parameters.}
\label{fig:axpby}
\end{figure}

As mentioned earlier, the {\tt axpby\_type} template parameter can be
used to reduce computation time for special values of $\alpha$ and
$\beta$ while maintaining a general interface.  \Fig{fig:axpby} shows
the speed and performance gain for single precision matrices of sizes
1-16 when $\alpha = 1$ and $\beta = 0$.  Depending on the matrix size,
the gains are between 10\% and 50\% over the code that works for
arbitrary $\alpha$ and $\beta$.  Hence, the added complexity due to
this extra template parameter is worth it.

\subsection{Speed for various data types and matrix sizes}

\begin{figure}
\centering
\includegraphics[page=4,scale=0.4]{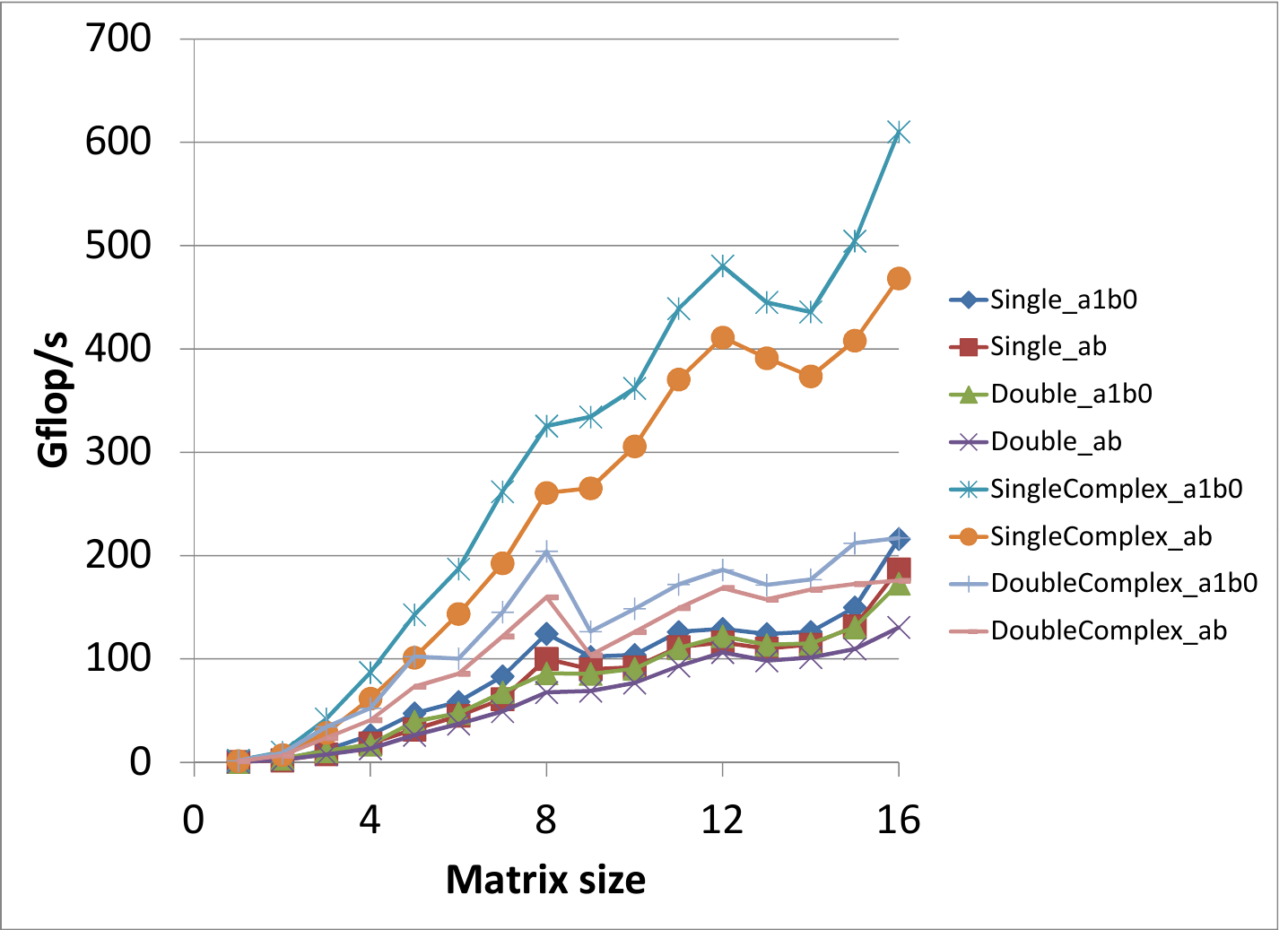}
\caption{Batched GEMM speed on NVIDIA Tesla K20c for various matrix sizes,
various data types, and when $\alpha = 1$ and $\beta = 0$ ({\tt a1b0})
or they are general values ({\tt ab}).}
\label{fig:all_types_speed}
\end{figure}

We now show the speeds achievable for four data types -- real and
complex types in single and double precision -- for matrix sizes 1-16.
We show results where $\alpha = 1$ and $\beta = 0$ in~\Fig{fig:all_types_speed}.
Naturally, using arbitrary $\alpha$ and $\beta$ will give a slightly lower
performance than what is shown. 
For matrices of size 16 and $\alpha = 1$ and $\beta = 0$ on NVIDIA Tesla K20c,
we reach these GFlop/s rates -- 216 for single real, 173 for double
real, 609 for single complex, and 217 for double complex.  For general
$\alpha$ and $\beta$, the speeds are lower by roughly 15\%-30\% for
matrices of size 16.

\subsection{Performance comparison with cuBLAS}

We now compare cuBLAS and our two kernels -- one which uses the second-order leading
dimension and one using pointers to pointers just like cuBLAS batched
interface.

As mentioned earlier, we have ignored memory allocation, deallocation, and transmission
overhead required in cuBLAS for working with pointers to pointers.  We have measured
only the time spent in calling GEMM. Similarly, we have ignored timing these
steps for out kernels, but naturally this time will be less than what is
required in cuBLAS interface.

\begin{figure}
\begin{center}
	\subfigure[Single precision real]
	{
		\includegraphics[scale=0.4,page=8]{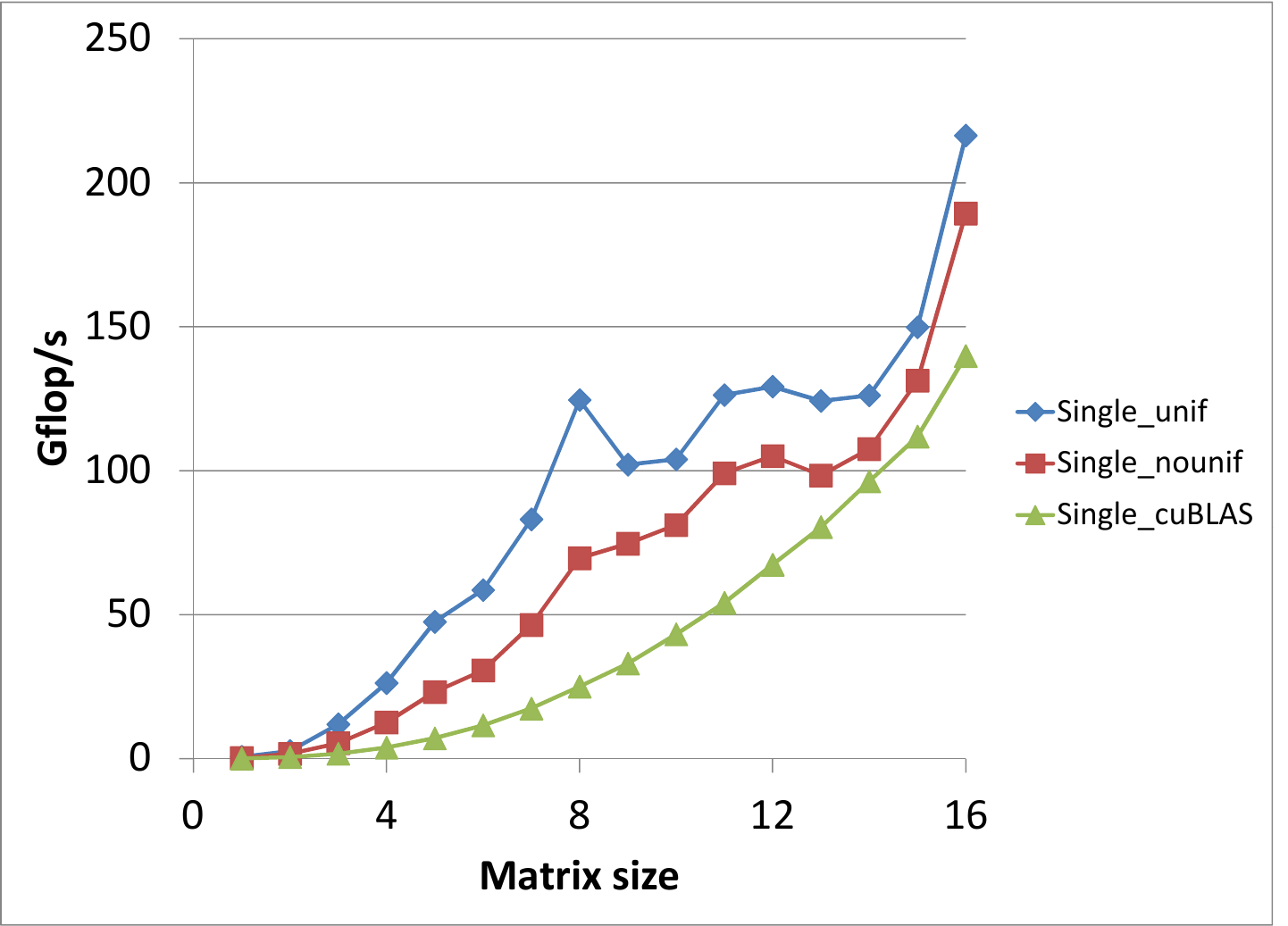}
	}\\
	\subfigure[Double precision real]
	{
		\includegraphics[scale=0.4,page=7]{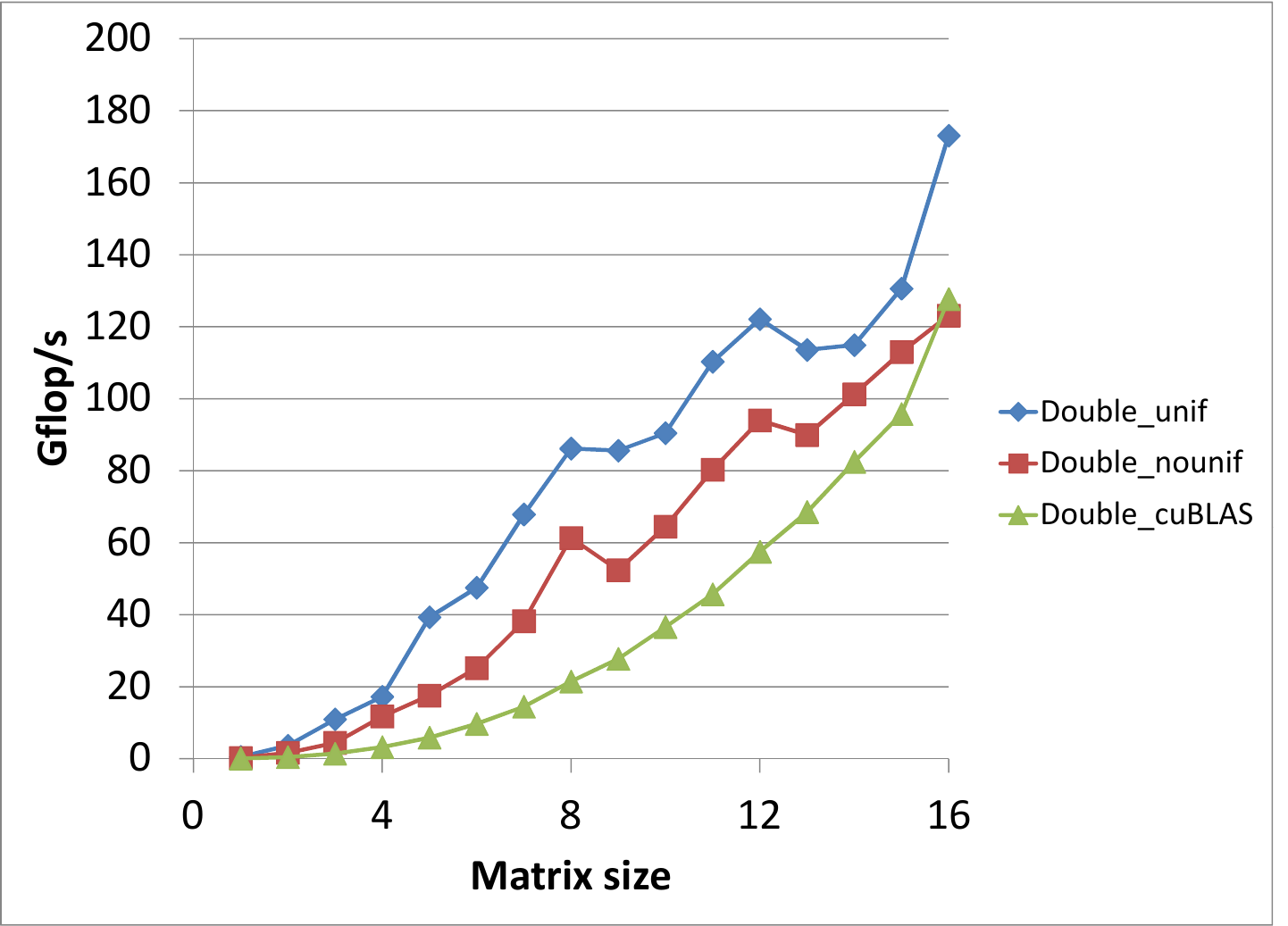}
	}
\caption{Comparison of the speed of our two kernels and cuBLAS batched
implementation for real scalars and $\alpha = 1$ and $\beta = 0$.
Relative improvement is shown in~\Fig{fig:rel_improve}.}
\label{fig:gemm_real_compare_cublas}
\end{center}
\end{figure}

\begin{figure}
\begin{center}
	\subfigure[Single precision complex]
	{
		\includegraphics[scale=0.4,page=6]{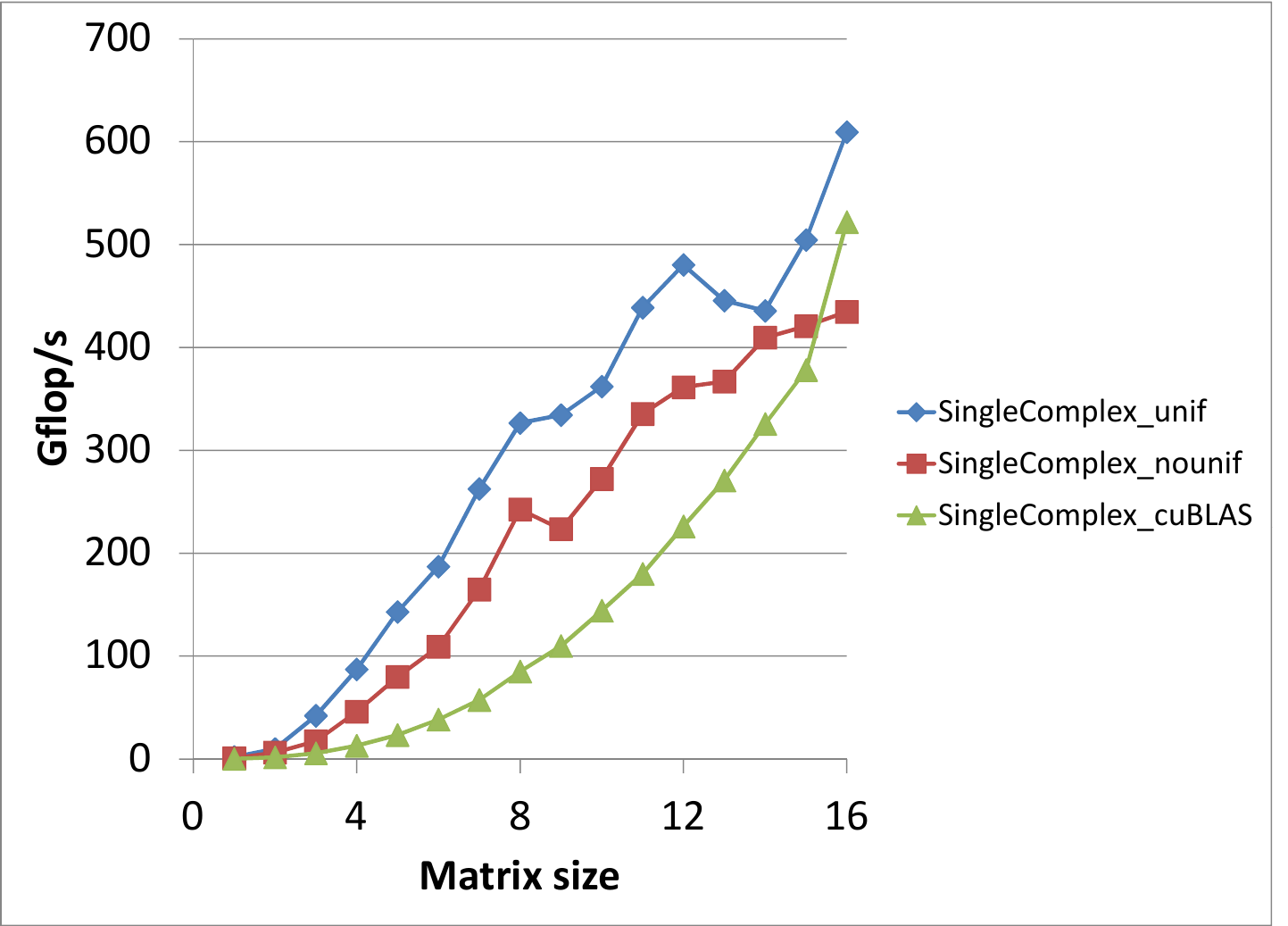}
	}\\
	\subfigure[Double precision complex]
	{
		\includegraphics[scale=0.4,page=5]{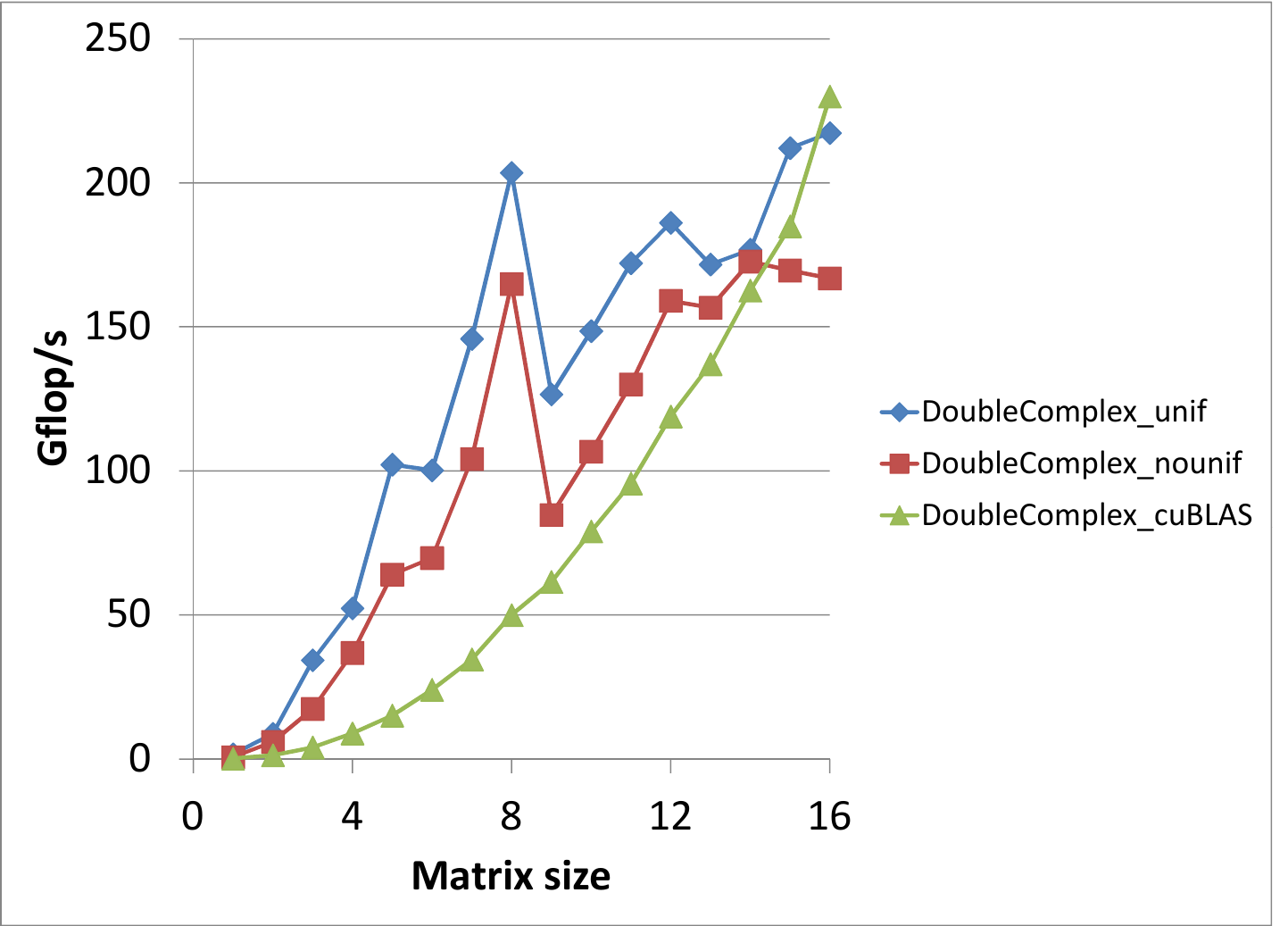}
	}
\caption{Comparison of the speed of our two kernels and cuBLAS batched
implementation for complex scalars and $\alpha = 1$ and $\beta = 0$.
Relative improvement is shown in~\Fig{fig:rel_improve}.}
\label{fig:gemm_complex_compare_cublas}
\end{center}
\end{figure}

See~\Fig{fig:gemm_real_compare_cublas} for the real case
and~\Fig{fig:gemm_complex_compare_cublas} for the complex case.
Both our kernels are noticeably faster than cuBLAS for
matrix sizes under 16 except for a handful of cases.  The second
order leading dimension based kernel, which is the one we intend to
use, is slower in just one case out of all by 5\%.

\begin{figure}
\centering
\includegraphics[page=9,scale=0.4]{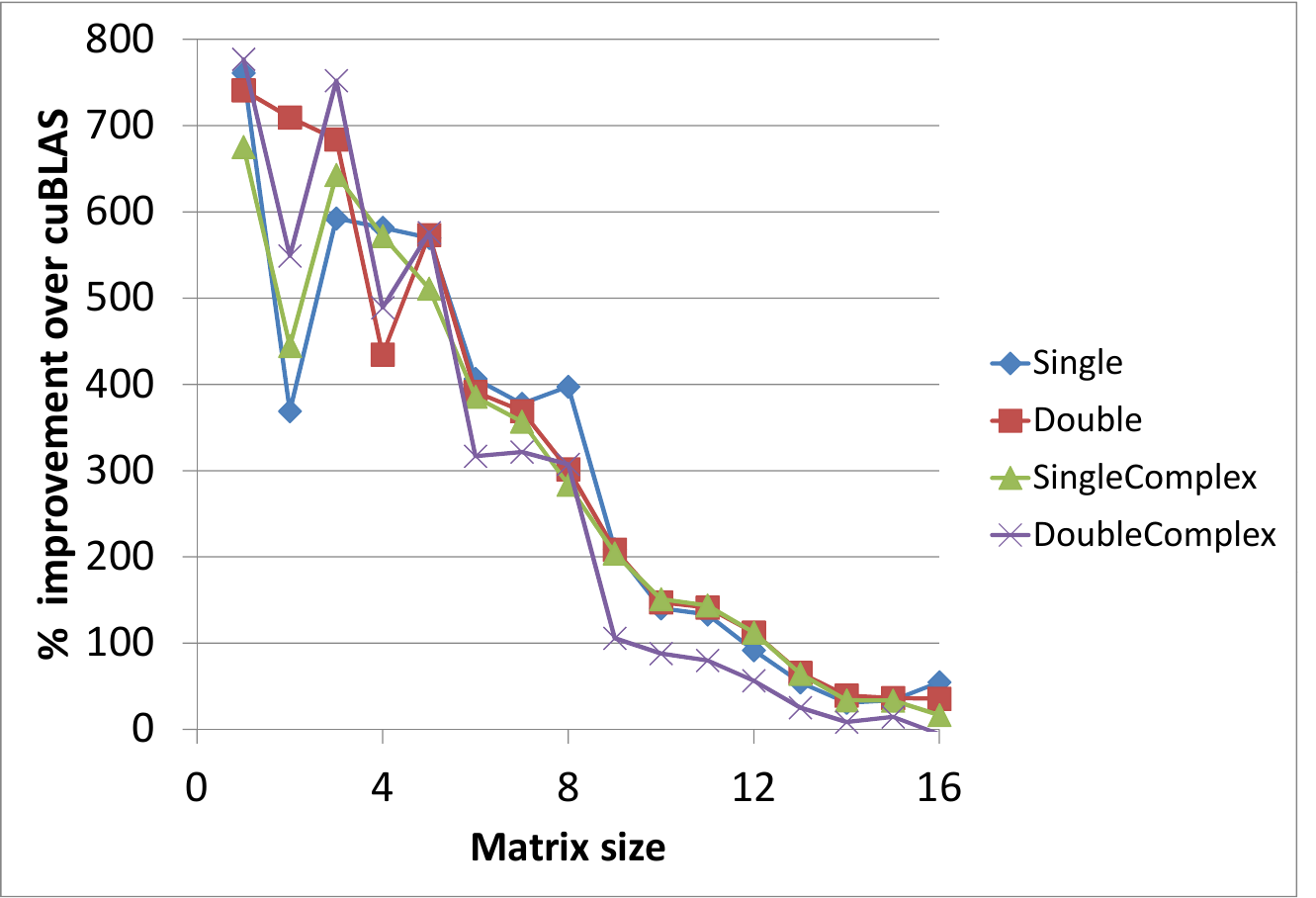}
\caption{Relative improvement over cuBLAS batched implementation
for various matrix sizes when our improved interface and
implementation is used.}
\label{fig:rel_improve}
\end{figure}

The results show that part of the improvement is due to our implementation
and part of it is due to the changes in interface.
\Fig{fig:rel_improve} shows the relative improvement for various
matrix sizes in the case of improved interface. It can be up to 600\%
for small sizes.  \Tab{tab:gflop} lists the speed we achieve when multiplying 100,000
independent matrix pairs of various sizes using the {\tt TGEMM\_multi\_uniform} kernel
on NVIDIA Tesla K20c.


\begin{table}
\begin{center}
\begin{tabular}{  r  r  r  r  r }
\hline
Size & Single-R & Double-R & Single-C &  Double-C \\ \hline
\hline                       
 1 &   1	&	 1	&	  2	&	2     \\ 
 2 &   3	&	 4	&	 10	&	9     \\ 
 3 &  12	&	11	&	 42	&	34    \\ 
 4 &  26	&	17	&	 87	&	52    \\ 
 5 &  48	&	39	&	143	&	102   \\ 
 6 &  59	&	48	&	187	&	100   \\ 
 7 &  83	&	68	&	263	&	146   \\ 
 8 &  125	&	86	&	327	&	204   \\ 
 9 &  102	&	86	&	334	&	127   \\ 
10 &  104	&	90	&	362	&	149   \\ 
11 &  126	&	110	&	439	&	172   \\ 
12 &  129	&	122	&	480	&	186   \\ 
13 &  124	&	114	&	446	&	172   \\ 
14 &  126	&	115	&	436	&	177   \\ 
15 &  150	&	131	&	504	&	212   \\ 
16 &  216	&	173	&	609	&	217   \\ \hline
\end{tabular}
\end{center}
\caption{The GFlop/s rates when multiplying 100,000 independent matrix pairs of various sizes using the {\tt
TGEMM\_multi\_uniform} kernel on NVIDIA Tesla K20c. The comparison with corresponding numbers in cuBLAS are made in
\Fig{fig:gemm_real_compare_cublas} and \Fig{fig:gemm_complex_compare_cublas}. The suffixes R and C
stand for real and complex, respectively.}
\label{tab:gflop}
\end{table}

\section{Discussion}

We have described a new interface and two improved implementations of
a batched GEMM routine.  Another interesting batched GEMM case is when
one matrix, either the one on the left or right, is fixed and one has to multiply it with a batch of other
matrices.  This can be treated by a non-batched GEMM implementation by
concatenating matrices but high performance may not be readily
possible because the non-batched GEMM might not be designed for such cases.  However, based on
our experience in this research, we believe that a special batched GEMM
implementation for this case will be much faster than using non-batched
GEMM.  Naturally, it will be faster than our case where both matrices vary.

The second leading dimension based interface presented here can be extended to other BLAS routines naturally,
for GPUs or multi-core CPUs.
As of now, we don't have data to quantify the importance of a batched GEMM implementation on GPU
in high performance of other GPU-based BLAS routines for small matrices.
However, just like the CPU case, we believe it will be crucial to formulate
other batched BLAS kernels in terms of batched GEMM.


\vspace{1cm}

\noindent \textbf{ \large{Acknowledgements}}

This work was partially supported by the US Department of Energy SBIR Grant
DE-SC0004439.

\bibliographystyle{elsarticle-num}
\bibliography{00_batch_gemm}

\end{document}